\documentclass[nofootinbib]{revtex4}
\usepackage{graphicx}
\usepackage{url}
\usepackage{latexsym}
\begin{document}
\input{epsf.sty}

\def\affilmrk#1{$^{#1}$}
\def\affilmk#1#2{$^{#1}$#2;}
\def\be{\begin{equation}}
\def\ee{\end{equation}}
\def\bea{\begin{eqnarray}}
\def\eea{\end{eqnarray}}


\centerline{\bf \Large Detecting the ambient neutralino dark
matter particles at accelerator}

\vspace{0.5cm}

\centerline{Tai-Fu Feng$^1$, Xue-Qian Li$^1$, Wen-Gan Ma$^3$,
Jian-Xiong Wang$^2$, and Gong-Bo Zhao$^{2}$}

\vspace{1cm}

1. Department of Physics, Nankai University, Tianjin 300071,
China.

2. Institute of High Energy Physics, Academy of Sciences of China,
Beijing 100049, China.

3. Department of Physics, University of Science and Technology of
China, Hefei 230026,China.

\vspace{0.5cm}
\begin{center}
\begin{minipage}{12cm}
In this work, we present a new strategy to investigate the
possibility of direct detection of the ambient neutralino matter at
accelerator. We calculate the cross sections for both elastic and
inelastic scattering processes of the dark matter particles with the
beam particles at $e^+e^-$ and hadron colliders.

\end{minipage}
\end{center}

\vspace{1cm}

The recent astronomical observation suggests that about $1/4$ of
the energy density in universe is contributed by dark matter which
does not participate in electromagnetic and strong interactions.
In fact, the dark matter particles only interact with other matter
particles via weak interaction and as well among themselves, thus
they are named as the WIMPs\cite{Hagiwara}.

Modern cosmology indicates the dark matter is "cold" rather than
"hot". The leading candidate of the cold dark matter particles in
the minimal supersymmetric standard model with R parity conserved
is neutralino $\widetilde{\chi}^{0}$.

The traditional method to detect the dark matter on Earth is to use
detectors with huge-volume. As the dark matter flux comes into
detector, an elastic scattering of neutralino
with proton in the detector: $\widetilde{\chi}^{0} + p \rightarrow
\widetilde{\chi}^{0} + p$, makes the proton recoiling. Since proton
is charged, a trajectory can be detected by sensitive electronics.
However, the kinetic energy of the neutralino is too small to cause
inelastic scattering such as $\widetilde{\chi}^{0} + p \rightarrow
\widetilde{\chi}^{+} + X$.

For such low energy, from other side, the cross section of the
elastic process $\widetilde{\chi}^{0} + p \rightarrow
\widetilde{\chi}^{0} + p $ is small and its theoretical prediction
on  its order of magnitude  is smaller than $10^{-6} \sim
10^{-7}$pb. The recent experiment of SOUDAN  sets an upper bound
of the cross section as $\sigma \leq 10^{-7}$pb. Because of the
small cross section and difficulty of the detection, one may turn
to search for other ways to directly detect the  dark matter flux.
Erede and Luk discussed a possibility of detecting the SUSY
particles in the cosmic rays at TEVATRON \cite{Luk}.

In this paper we study a new mechanism to detect cold dark matter by
using the accelarator beam particles to collide with the ambient
neutralino dark matter particles. Obviously, direct detection of the
dark matter particles via inelastic scattering is very beneficial
because the products of the scattering can involve a heavy charged
SUSY particle whose trajectory would be clear thus easy to detect.
Moveover, in the elastic scattering of $\widetilde{\chi}^{0} + p
\rightarrow \widetilde{\chi}^{0} + p$, the background is hard to be
fully eliminated, namely the contamination from the background
proton or other charged particles makes  the experimental
identification of $\widetilde{\chi}^{0}$  even more difficult. The
inelastic scattering does not suffer from such problems.

In this work we investigate a possibility of direct detection of the
dark matter flux at accelerators via elastic and inelastic
processes. The accelerators available at present or will be
available in the near future are LEPII, TEVATRON, LHC and ILC. The
concrete inelastic processes are that one uses the beam of extra
high energy at accelerator to bombard on the dark matter particles
which come into the detector and even the beam tube and then charged
SUSY particles $\widetilde{\chi}^{+}$ are produced  via the
scattering, i.e. the processes such as $\widetilde{\chi}^{0} + e^{-}
\rightarrow\widetilde{\chi}^{-} + \nu_e$, $\widetilde{\chi}^{0} +
e^{-} \rightarrow \widetilde{\chi}^{-} +\gamma$ or
$\widetilde{\chi}^{0} + p \rightarrow \widetilde{\chi}^{+} + X$ etc.

The detection rate is
\begin{equation}
\label{number}
 N=\rho_{dm}  \rho_{beam} |\upsilon_{rel}| \sigma s l  t \alpha
\end{equation}
where $\rho_{dm}$ is the dark matter density in the ambient space
of our Earth, $\rho_{beam}$ is the flux of $e^{-}$ or proton at
the accelerator, $\mathbf{v}_{rel}$ is the relative velocity of
the beam and dark matter flux, but as the previous study
indicates, the velocity of dark matter cannot exceed 1000km/s,
thus compared to the velocity of the beam particle which is very
close to the speed of light $c$, one can treat the dark matter
particle to be at rest in the laboratory frame as $
|\mathbf{v}_{rel} | \sim c$. $\sigma$ is the scattering cross
section which we evaluate in this work. $s$ is the cross section
of the beam, $l$ is the length of the detector and $t$ is the time
duration of the experiment, and $\alpha$ is the detection
efficiency.

In the regular non-accelerator experiments, this equation still
holds, but $\rho_{beam}$ is replaced by the density of proton in
the detector, $sl=v$ is the volume of the detector and
$|\mathbf{v}_{rel} |$ is the velocity of the dark matter particles
in the lab frame. In that case, the volume of detectors can be
very large, but as aforementioned, in this case $\sigma$ is small
and detection efficiency is low. By contrary, in the accelerator
case, $\sigma$ can be enhanced by $6 \sim 7$ orders, the detection
efficiency can be apparently improved, but the effective volume
$sl$ is much smaller than that of detectors used for
non-accelerator experiments. The purpose of this work is that if
we can swap the loss of the effective volume  by the advantages of
high cross section and detection efficiency. The key point is how
much the cross section is enhanced and the decisive factor is the
effective volume provided by the available accelerators and
detectors and the dark matter flux.

In this work, we have re-done the calculations on the cross sections
of $\tilde\chi^0+e^-\rightarrow \tilde\chi^-+\nu$ and
$\tilde\chi^0+p\rightarrow \tilde\chi^++X$ in light of \cite{Feng},
by the results, we obtain the corresponding detection rates. We take
several groups of the parameters for the minimal SUSY model with the
mSUGRA scheme. The parameters are determined based on the present
collider data and are called the benchmark parameters by the
Snowmass.

\noindent 1. The elastic scattering of neutralino with the
electron beam.

The process under consideration is
\begin{equation}
\tilde{\chi}_1^0+e^-\rightarrow \tilde{\chi}^0_{1}+e^-,\;\;\;\;
{\nu}_e+e^-,
\end{equation}
where we assume that the lightest supersymmetric particle is
neutralino $\tilde{\chi}^0_{1}$.

The traditional method for detecting the dark matter flux which is
composed of weakly interacting $\tilde{\chi}^0_{1}$'s, is to let
them collide with the nucleons or electrons in the detector and
measure the trajectories of the recoiled charged SM particles. The
available energies for the elastic scattering is very low and
cross section is small. In accelerator, as one uses the beam
particle (electron or positron) to bombard on the dark matter
flux, and one can observe that some of the projectile electrons
decline from the beam direction, so the signal is clear, and the
energies are much higher, resultant cross section may be
increased.  With various electron beam energies, we obtain the
cross sections which are tabulated in Table 1.

\vspace{0.3cm}

\begin{center}
\begin{tabular}{cc}
\hline
\hline
$E_n$ (GeV) & $\sigma_{total}\;(pbar)$ \\
\hline
0.30E+01 & 0.44 E-2\\

0.50E+01 & 0.11 E-01\\

0.10E+03 & 0.23 E+00\\

0.25E+03 & 0.46 E+01\\

0.75E+03 & 0.30E+01\\

0.20E+04 & 0.10 E+01\\
\hline \hline
\end{tabular}
\end{center}

\vspace{0.3cm}

\centerline{Table 1. The cross section for elastic process
$\tilde{\chi}_1^0+e^-\rightarrow \tilde{\chi}^0_{1}+e^-$ with
various beam energies.}

\vspace{0.3cm}

Meanwhile, a background may contaminate the situation. The
observation is based on measuring the electrons scattered from the
SUSY dark matter particles in $e^-+\tilde{\chi}_1^0 \rightarrow
e^-+\tilde{\chi}_1^0$ and there is a background from the electrons
scattered from nucleons of the remnant atmosphere in the
vacuumized tunnel, $e^-+n\rightarrow e^-+n$. At lower energies,
the cross section of scattering can be easily computed and the
amplitude is
\begin{equation}
{\cal M}={G_F\over 2\sqrt 2}\bar n\gamma_{\mu}[(-1+{4\over
3}\sin^2\theta_W)+\gamma_5]n \bar
e\gamma^{\mu}[(-1+4\sin^2\theta_W)+\gamma_5]e.
\end{equation}
Then we can obtain the cross section. At the same length, the
background events are at least 1000 times larger  than the
expected events at $1.0315 10^{-6} pa$.

Recently, Hisano et al.\cite{Hisano}, also suggest to measure the
number of electron recoil events by neutralino at accelerator.
They conclude that if very high current beam is available, the
dark matter wind can be observed.\\

2. The inelastic cases.

As discussed above, the elastic scattering may suffer from
mis-identification of the signal from the background. We would
turn to study if one can measure the dark matter flux via
inelastic scattering between the projectile and the neutralinos.

(a) In the $e^+e^-$ colliders.

If the kinematics is permissive, several inelastic reactions such
as $e^{-}+\tilde{\chi}_{1}^{0} \rightarrow \tilde{\nu}_{e} +
W^{-}(H^{-}_{1})$, $e^{-}+\tilde{\chi}_{1}^{0} \rightarrow
\tilde{e}_{i}^{-} + Z^{0}(H^{0}, A^{0})$ $(i=1$, $2)$ etc. can
occur. However, without losing generality, we suppose that
$\tilde{\chi}_1^-$ is the lightest charged leptonic SUSY particle,
so that at the moment we consider only the inelastic channel
$e^{-}+\tilde{\chi}_1^0 \rightarrow \tilde{\chi}_1^- + \nu_{e}$.
The expressions of the corresponding amplitudes for the reaction
were given in our earlier work \cite{Feng}.

The results somehow depend on the mass difference of
$\tilde{\chi}_{1}^{0}$ and $\tilde{\chi}_1^-$.

As an example, we would like to investigate a special case. In
1972, a peculiar event of heavy cosmic ray particle was observed
in the cloudy chamber of the Yunan Cosmic Ray Station (YCRS)
\cite{Wuli}. Recently, the event was re-analyzed \cite{Ho} and it
is identified as that a heavy neutral particle $C^0$ came in and
bombarded on a proton to produce a heavy charged particle $C^+$ as
well as a proton and $\pi^-$. Their analysis confirmed that the
mass of the heavy neutral cosmic ray particle $C^0$ is greater
than 43 GeV and the mass difference
$$\Delta M=M_{C^+}-M_{C^0}<0.270\;{\rm GeV}.$$

If taking this result seriously, one would be tempted to conclude
that the coming neutral $C^0$ is a SUSY dark matter particle
$\tilde{\chi}_1^0$ and the produced heavy charged particle is
$\tilde{\chi}_1^+$ accordingly.

The cross sections are listed in Table 2.

\vspace{0.3cm}

\begin{center}
\begin{tabular}{cc}
\hline
\hline
$E_n$ (GeV) & $\sigma_{total}\;(pbar)$ \\
\hline
0.30E+01 & 0.40 E+00\\

0.50E+01 & 0.11 E+01\\

0.10E+03 & 0.70 E+02\\

0.25E+03 & 0.13 E+03\\

0.75E+03 & 0.11 E+03\\

0.20E+04 & 0.14 E+03\\
\hline
 \hline
\end{tabular}
\end{center}
\vspace{0.3cm} \centerline{Table 2. The cross section for
inelastic process  $\tilde{\chi}_1^0+e^-\rightarrow
\tilde{\chi}_1^-+e^-$ with various beam energies.}

\vspace{0.3cm}

The number density of the projectile beam is
\begin{equation}
\rho_{beam}={No.of particles\;per\;bunch\over bunch\;length\cdot
S},
\end{equation}
the total event number one may expect to observe is calculable. As
$\rho_{DM}$ is 0.3 GeV/cm$^3$, with the optimal parameters which
we can find from the available accelerators in the world
\cite{Data}, we achieve that $$N=8\times 10^{-5}\; events/year,$$
for $l=1$ m.\\

(b) In the hadron colliders.

It is natural to expect that at the hadron colliders, the
situation might be remedied, because the available beam energies
are much larger and the number of partons which may contribute to
the total inclusive cross sections would be greatly increased. The
concerned sub-processes are
\begin{eqnarray}
u+\tilde{\chi}_1^0\rightarrow d+\tilde{\chi}_1^+ \nonumber\\
d+\tilde{\chi}_1^0\rightarrow u+\tilde{\chi}_1^- \nonumber\\
s+\tilde{\chi}_1^0\rightarrow c+\tilde{\chi}_1^- \nonumber\\
c+\tilde{\chi}_1^0\rightarrow s+\tilde{\chi}_1^+ \nonumber\\
b+\tilde{\chi}_1^0\rightarrow t+\tilde{\chi}_1^- \nonumber\\
\bar u+\tilde{\chi}_1^0\rightarrow \bar d+\tilde{\chi}_1^- \nonumber\\
\bar d+\tilde{\chi}_1^0\rightarrow \bar u+\tilde{\chi}_1^+ \nonumber\\
\bar s+\tilde{\chi}_1^0\rightarrow \bar c+\tilde{\chi}_1^+ \nonumber\\
\bar c+\tilde{\chi}_1^0\rightarrow \bar s+\tilde{\chi}_1^- \nonumber\\
\bar b+\tilde{\chi}_1^0\rightarrow \bar t+\tilde{\chi}_1^+.
\end{eqnarray}
The effective lagrangian can be found in literature \cite{Haber}.

In terms of the parton distribution function \cite{Martin}, we use
the FDC program \cite{Wang} to calculate the cross sections for
$\tilde{\chi}_1^0+p\rightarrow \tilde{\chi}_1^{\pm}+X$. The key
point is the various SUSY breaking mechanisms which may result in
different values for the cross sections. Let us list the possible
parameter space with the two typical breaking mechanisms in Table
3.

\vspace{0.3cm}
\begin{center}
\begin{tabular}{lllllll}
\hline \hline

Symbols  & schemes & $m_0$ (GeV) & $m_{1/2}$ (GeV) & $A_0$  & $\tan\beta$ & $sign\; of \;\mu$ \\
\hline
1 & typical values & 100 & $-100$ &0 &10 & +\\

2  & region of focus  & 1450 & 300 & 0  & 10 & + \\

3  & region of annihilation  & 90 & 400 & 0  & 10 & + \\

4  & large $\tan\beta$  & 400 & 300 & 0  & 50 & + \\

5  & light $\tilde t$  & 150 & 300 & -1000  & 5 & + \\

6  & non-universal gluino masses  & 150 & 300 & 0  & 10 & + \\
\hline

 & GMSB & $\Lambda$ & $M_{mess}$ & $N_{mess}$ & $\tan\beta$ & $sign\; of
 \;\mu$\\
\hline
7 & $NLSP=\tilde\tau_1$ & 40,000 & 80,000 & 3 & 15 & +\\

8 & $NLSP=\tilde\chi^0_1$ & 100,000 & 200,000 & 1 & 15 & +\\
\hline
 & AMSB &  $m_0$ & $m_{3/2}$ & & $\tan\beta$ & $sign\; of
 \;\mu$\\
\hline 9 & small $m_{\tilde \chi_1^{\pm}}-m_{\tilde\chi_1^0}$ & 400
&
60,000 & & 10 & + \\
\hline \hline
\end{tabular}
\end{center}

\vspace{0.3cm}

\centerline{Table 3. The parameters adopted for the later
numerical computations}

\vspace{0.3cm}

In Table 3, we only list the necessary parameters for the SUSY
sector, and for the SM sector all the parameters can be found in
the data book \cite{Data}. In the table, we include two
SUSY-breaking scenarios and the corresponding SNOW-MASS 2001
benchmark points suggested by the SNOW-MASS working group.

With the parameter as inputs, we calculate the cross section of
$\tilde{\chi}_1^0+p\rightarrow \tilde{\chi}_1^{\pm}+X$ and the
numerical results are tabulated in Table 4.

\vspace{0.3cm}

\begin{center}
\begin{tabular}{ll}
\hline
\hline No. & Total cross section $\sigma$ (pb) \\
\hline 0 & 0.73E+03\\
 1 & 0.13E+02\\
 2 & 0.12E+01\\
 3 & 0.56E+01\\
 4 & 0.75E+01\\
 5 & 0.93E+02\\
 6 & 0.76E+01\\
 7 & 0.87E+01\\
 9 & 0.22E+01\\
\hline \hline
\end{tabular}
\end{center}

\vspace{0.3cm}

\centerline{Table 4. The cross sections of the inelastic
scattering $\tilde{\chi}_1^0+p\rightarrow
\tilde{\chi}_1^{\pm}+X$.}

\vspace{0.3cm}

In Table 4, the first row (No.0) corresponds to the special case
where we adopt the parameters determined by the data of the Yunan
observatory \cite{Ho} for a comparison. Namely there we set
$m_{1/2}=250$ GeV, $A_0=-100,\;\tan\beta=10,\;\mu>0$ and
$m_{\tilde\chi_1^0}=43$ GeV, $m_{\tilde\chi_1^-}=43.27$ GeV. It is
obvious that the obtained cross section with this group of
parameters is larger than the typical values by one order.

For LHC, the beam energy is about 7000 GeV, if we take the maximum
cross section to be $\sigma\sim 727$ pb (No.0 in Table 4), and the
ideal situation with the detection efficiency $\eta\sim 1$, one can
expect $$N=1.4\times 10^{-5}\; events / year.$$

Moreover, it is impossible to use the elastic scattering to
measure the dark matter flux, because in that case the recoil
trajectory of the beam particles would be drowned in an incredible
background.\\

As discussed at beginning, the accelerator experiments may provide
higher energies which can be used to bombard on the dark matter
particles coming to our earth. No matter in the elastic scattering
at $e^+e^-$ colliders or inelastic scattering at both $e^+e^-$ and
hadron colliders, one can expect clearer signals. Because in the
traditional method where the dark matter particles scatter with
the proton or electron in the detector and the trajectory of
proton or electron recoils are measured. The kinetic energy of the
dark matter particle is small (about 0.3 to 0.6 MeV) and the
corresponding cross section is small too. With this small
recoiling kinetic energy, the trajectory of charged particle
(proton and electron) is not clear. However, the detector can be
made to possess a large volume and the number of particles which
may interact with the dark matter particles is large and it can
compensate the small reaction cross sections.

On other side, the elastic or inelastic scattering processes of the
SM particles and the dark matter particles at accelerators of very
high energies, can result in larger cross sections and provide clear
signals which almost cannot be misidentified with the background.
However, the volume of the detector which would be the accelerator
itself at our proposal, is very limited. Generally, the size of the
beam bunch is of order 100 $mm^3$ and each bunch may contain at most
$10^{12}$ particles.

Therefore, our conclusion is that unless one can greatly increase
the luminosity or the detector length (it seems impossible to
enlarge the beam radius) by at least 6 to 7 orders, it is
impossible to obtain any data which has substantial significance.

However, it seems not to be the end of the story, because the
present knowledge on the distribution of dark matter in the space
tells us that the density at the ambient space of our earth might
be much larger than the universe-averaged value. If it happens
that the dark matter density at the ambient space of our earth is
several orders larger than the universe-averaged value which is
used in our above computations, the flux may be possible to be
observed. It is also noted that because the signal is clean, the
requirement on the detector quality is not as rigorous as that for
regular experiments. Thus if the detector can be made to be longer
than 1 m, say, 100 m, then we would have a larger chance to
directly observe the dark matter flux and the idea is very
encouraging. The observation may be a nice complementarity to the
direct search for SUSY particles by producing them at
accelerators.\\

{\bf Acknowledgements:} This work is partially supported by the
National Natural Science Foundation of China and the Special
Research Fund for the Ph.D programs of Chinese Universities. The
authors would like to thank Dr. X.M. Zhang for very helpful
discussions, in fact, he contributes substantially from beginning
of the work to the final version. We would like to thank Dr. J.
Feng who pointed out the mistakes in our earlier work where we
greatly overestimated the density of dark matter in our ambient
space. We benefit greatly from the talks by Dr. S.F. Su given at a
workshop at CCAST from Dec. 20 to
23 of 2005.\\

\end{document}